# INTRODUCING ASTROGEN: THE ASTRONOMY GENEALOGY PROJECT

**Joseph S. Tenn**

*Sonoma State University, Rohnert Park, CA 94928, USA.*
Email: joe.tenn@sonoma.edu

**Abstract:** The Astronomy Genealogy Project ("AstroGen"), a project of the Historical Astronomy Division of the American Astronomical Society (AAS), will soon appear on the AAS website. Ultimately, it will list the world's astronomers with their highest degrees, theses for those who wrote them, academic advisors (supervisors), universities, and links to the astronomers or their obituaries, their theses when on-line, and more. At present the AstroGen team is working on those who earned doctorates with astronomy-related theses. We show what can be learned already, with just ten countries essentially completed.

**Keywords:** Academic genealogy, astronomers, Ph.D. theses, dissertations

## 1 INTRODUCTION

AstroGen is coming. The Astronomy Genealogy Project will soon appear on the website of the American Astronomical Society (AstroGen: https://astrogen.aas.org/). Under construction since early 2013, the project will list the world's doctoral theses (dissertations) on astronomy-related topics, along with information about the theses and their authors.

The original goal was to emulate, and possibly improve upon, the highly successful Mathematics Genealogy Project (MGP: http://www.genealogy.ams.org/), which has been underway since 1996 and currently holds information about more than 200,000 'mathematicians'. This number includes more than a thousand whose 'math subject area' is listed as 'astronomy and astrophysics' and several thousand classified in at least eight fields of physics.

Note that in academic genealogy one's parent is one's thesis advisor (also known as supervisor, *directeur, Doktorvater, promotor* …). Academic genealogy sites allow a scholar to trace his or her academic ancestors, and many find this enjoyable. I found my academic grandfather listed in the MGP, so I entered my academic father and myself, even though our degrees are in physics, and now a visitor to the MGP can trace my academic ancestry back 29 generations to the year 1360. Of course most of the early generations lacked doctorates, and the information about them is sketchy. Before the seventeenth century nearly all the degrees were in medicine, theology, or law, and many were not doctorates. Many scholars did not even take degrees.

The modern doctorate, usually called a Doctor of Philosophy, or Ph.D., began in Germany in the early nineteenth century. It gradually spread to most countries over the next century, although it did not become popular in some places, notably the United Kingdom, until after World War II. The first granted in the ten countries we have studied (listed in Section 3) went to Arthur Williams Wright (Figure 1), who became the first person outside Europe to earn a Ph.D. in science and one of the first three Ph.D.s in any subject in the United States. His thesis, *Having Given the Velocity and Direction of Motion of a Meteor on Entering the Atmosphere of the Earth, to Determine its Orbit about the Sun, Taking into Account the Attraction of Both these Bodies,* was submitted to Yale College in 1861. A man of many talents, Wright earned a law degree, tutored Latin, and ended up as a professor of physics at Yale, where he made some of the earliest experiments with X-rays.

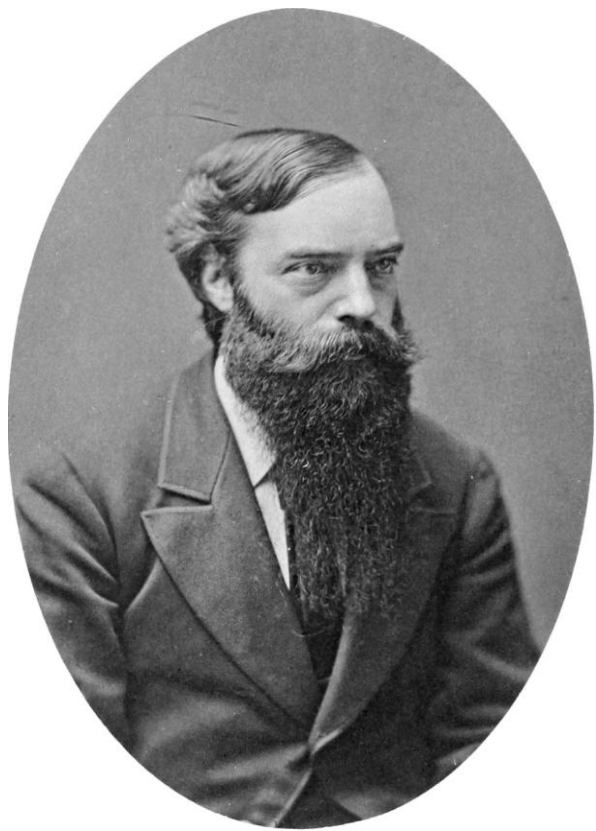

Figure 1: Arthur Williams Wright (1836−1915). Photographer unknown (after Kingsley, 1879: 431).

## 2 WHAT ARE WE DOING?

We have been filling in a spreadsheet to be converted to a proper database by the IT experts who work for the American Astronomical Society. It has 30 columns for each dissertation, and almost every one of them causes questions and conflicts and forces us to make arbitrary decisions.

The biggest question is whom to include. For now we are including every astronomy-related thesis. We use this word because *thesis* is widely understood worldwide, although the proper term in our own country, the United States, is *dissertation* for the Doctorate, while *thesis* is used for the Master's degree. Some countries follow the reverse convention. We define 'astronomy-related' to include the scientific study of anything that is or comes from outside the Earth, and the development of tools to facilitate such study. We find that such theses are earned in a variety of academic departments—Astronomy and Physics of course, but also Aerospace Engineering, Chemistry, Computer Science, Earth Science, Electrical Engineering, Geology, Mathematics, Mechanical Engineering, Meteorology, Space Science, and others. In seeking such theses we often find it convenient to search for 'astronomy or astrophysics or cosmology or planetary science'. Note that we exclude theses on ethnoastronomy, archaeo-astronomy, history of astronomy, and education in astronomy, even though degrees on such topics are occasionally awarded by Astronomy Departments.

There are grey areas. How much of cosmology should we include? Observational cosmology for sure, but what about theoretical theses? We have included most while trying to exclude those that are so purely theoretical that they show no connection with observations (e.g., brane theory, string theory), but it may be impossible to be completely consistent. We have included searches for dark matter in nature, but excluded attempts to make it in accelerators. Another big problem area is near-Earth geophysics. Sometimes we have to resort to the arbitrary definition that space begins 100 km above the ground. We have included the study of the ionosphere and beyond. We exclude studies of the interior of the Earth unless they compare it with some other planet or planets. The inclusion of the development of tools for astronomy is another complicated area. Design of a new optical telescope or spectrometer? Yes. Lightning protection and radio frequency interference mitigation for a new radio telescope? Yes, for now, but we are not sure. Design of a rover for planetary exploration? No, for now, but we are uncertain.

There are many other places where we have made somewhat arbitrary decisions. Some may yet be changed.

## 3 WHAT HAVE WE ACCOMPLISHED SO FAR?

As of November 2016 we have entered more than 20,000 theses, including nearly complete coverage of ten countries—Australia, Canada, Chile, Ireland, the Netherlands, New Zealand, South Africa, Sweden, the United Kingdom and the United States. We have started with theses in the language we know best. (The last Dutch thesis on an astronomy-related subject in a language other than English was submitted in 1962. Of the 37 Chilean theses, 31 are in English. All post-1800 theses from Sweden that we have found are in English.)

Here are the items we have attempted to record, with some of the questions that have arisen for each.

### 3.1 Name

People change their names for a variety of reasons. For example, women in the Western world have long changed their surnames on marriage, and sometimes again on divorce. While this is becoming less common, it still leads to difficulties in identifying some. Astronomers from east Asia, where the family name comes first, study in the West, where they reverse name order for the thesis and a few publications. Some then return to their home-lands and change the order of their names back. Scholars in Spanish-speaking countries use their full, formal names, consisting of given names followed by father's surname and then mother's surname, on their theses, but many then omit the mother's surname on their websites and publications (and just to make it more confusing, some combine father's and mother's names with hyphens). And not a few immigrants change their names to make them easier for residents of their adopted country to remember and pronounce. A few change their given names because of changing genders. Some have other personal reasons. We have tried to put the last-used name at the top of this category, but to include other names for those who wish to trace publications and careers. There is also the problem of two or more astronomers with identical names. This is especially common among those of Chinese or Korean origin. The practice of including only initials and surname on a thesis, while waning, is still a pernicious one from our perspective. 'Y. Wang' publishes more than ten scientific papers per day (Butler, 2012). Of course there are many Y. Wangs (does anyone know how many?), which is why it is important that scientists acquire and use identification numbers such as those of ORCID (2016).

### 3.2 Years of Birth and Death

We have recorded these when we have come across them, but we do not intend to make birth years public for living persons.

### 3.3 University Granting the Degree

Universities change their names; they merge; they split. Some use different names in different languages. This becomes complicated. On a person's page, we will have a link from the name of the university granting the degree (at the time, but in English) to a page for that university. There we will give the other names used by the university, including those in its own language(s). We will also give the current name, the country where the university is now located, and a link to the university's website if it exists. (A small number of universities have ceased to exist.)

### 3.4 Name of the Degree

How do we translate doctorates in other languages? At present we are using 'Ph.D.' for nearly all earned doctorates, even though they may be called 'Doctor of Physical Science', 'Doctor of Astronomy', or something else. This appears to be the custom for those who earn doctorates nowadays in countries where a different title is used. Then there is 'D.Sc.' At one time several American universities awarded it interchangeably with the Ph.D. This was done at the Massachusetts Institute of Technology as recently as 1992. Many universities, especially in Australia, award a D.Sc. as an honorary degree, but they request a 'thesis', which consists of a bundle of previously-published papers. This has served a valuable purpose in recognizing distinguished senior scientists, including several who were too busy founding radio astronomy after WWII to bother earning a doctorate. We are including those awarded this degree with submission of a 'thesis' if they did not have a previous doctorate. If a D.Sc. was awarded with no thesis, we exclude it, as it is usually the kind of honorary degree awarded to donors.

### 3.5 Year of the Degree

We have tried to use the year the degree was awarded, but that is not always available. We find theses in libraries, and librarians are more interested in copyright dates. The date on the thesis is often the date the thesis was submitted or defended. If this is late in the year, the degree may well have been awarded the following year. In a few cases we have found that the degree was awarded two or more years after the thesis was defended, presumably because some other degree requirement was not met. We expect it to be impossible to please everyone with the years we have listed.

### 3.6 Thesis Title

We intend to include the original title and an English translation for those in other languages. This assumes we can find volunteer translators.

### 3.7 Advisors and Mentors

Until recent decades, most thesis research was directed by a single advisor (supervisor). A few students had two. Now it is not uncommon for a student to have three or even four advisors, and it is quite common for a new Ph.D. to thank many, many scientists for being very helpful in the research that led to the thesis. We have been including mentors as a separate category, restricting the title of mentor to those who are called unofficial or *de facto* advisors in the acknowledgement pages of the thesis. Yes, we have read thousands of such pages. It is sometimes difficult to separate the official advisors from the mentors. We have obtained the names of the advisors of more than 82% of the Ph.D.s we have recorded in the ten countries mentioned above, in a few cases by examining the theses in libraries, but in the great majority from on-line sources.

### 3.8 A Link to the Thesis if it is On-line

Some readers may be surprised at how many theses are on-line. Of the 18,923 theses we have recorded for doctorates awarded in the ten countries from 1861 through November 2016, 37% are freely available to everyone on the world wide web, and another 28% are on ProQuest (2016), a database to which a great many academic libraries subscribe. ProQuest is the successor to University Microfilms, which microfilmed most American theses for many decades. It now includes other countries as well, and most theses have been digitized. Starting at various dates since the mid-1990s, most universities have required that all theses be submitted in electronic format. It is quite possible that many American theses never see paper, while in the Netherlands they are printed and bound with handsome covers despite the fact that they are submitted electronically. Some universities make all or nearly all their theses freely available on their websites, while others limit viewing to members of their own campus communities. Of course the author, as copyright holder, has to give permission, and some authors embargo their theses for a year or two or three.

Table 1: Progress through November 2016. The second column is the current population in millions. The third column is the number of institutions in a country that have awarded two or more doctorates with astronomy-related theses. (In this column we ignore institutions that have awarded just one, but their output is included in column 5.) The number of degrees is from 1861 through late 2016, but is more up-to-date for some universities and countries than others. The last two columns are the year of the first modern doctorate and the median year for production of doctorates. Populations are from Wikipedia (2016). Degrees awarded by two universities for one thesis are counted only once. Subtraction of duplicates is done within country totals where the two universities are in the same country, but one degree was awarded by universities in two different countries and the duplication was subtracted in the total, which is why the number of degrees for all countries is one less than the sum of the column above it.

| Country | Population (million) | No. of Instns | Instns/ pop | Number of Doctorates | Degrees/ pop | Degrees/ Instn | First Year | Median Year |
|---|---|---|---|---|---|---|---|---|
| Australia | 24.2 | 17 | 0.70 | 702 | 29 | 41 | 1953 | 2001 |
| Canada | 36.5 | 23 | 0.63 | 829 | 23 | 36 | 1926 | 2001 |
| Chile | 18.2 | 3 | 0.16 | 37 | 2 | 12 | 2004 | 2012 |
| Ireland | 4.8 | 6 | 1.25 | 103 | 21 | 17 | 1967 | 2010 |
| Netherlands | 17.0 | 8 | 0.47 | 939 | 55 | 117 | 1863 | 1999 |
| New Zealand | 4.7 | 4 | 0.85 | 84 | 18 | 21 | 1957 | 2004 |
| South Africa | 55.7 | 7 | 0.13 | 97 | 2 | 14 | 1972 | 2009 |
| Sweden | 9.9 | 8 | 0.81 | 363 | 37 | 45 | 1853 | 2005 |
| United Kingdom | 65.1 | 35 | 0.54 | 2890 | 44 | 83 | 1904 | 2002 |
| United States | 324.3 | 153 | 0.47 | 12880 | 40 | 84 | 1861 | 1997 |
| **All ten countries** | **560.4** | **264** | **0.47** | **18923** | **34** | **72** | **1861** | **1999** |
| California | 39.1 | 11 | 0.28 | 2442 | 62 | 222 | | |

### 3.9 A Link to the Author's Web Page or Obituary

This may be foolish, as web pages change so frequently, especially in early careers. Yet we are trying it, and we hope that astronomy graduates will send us updates once we are on-line.

### 4 WHAT HAVE WE LEARNED?

As we look over the results we discover that AstroGen can be used for much more than tracing one's academic ancestry. It can be valuable to historians and sociologists of science who will be able to compare universities, countries, and eras. For examples of such research, see Gargiulo et al. (2016) and references cited therein. (There are other uses as well. A significant use of the MGP has been by editors wishing to avoid sending papers for review to the advisors or students of authors.)

I have compiled some information for the ten countries which are nearly complete. It is of interest to compare their astronomy-related degree production with their populations and the number of universities granting these degrees. Table 1 shows this for the ten countries, plus one subdivision, the U.S. state of California.

The number of universities that have granted two or more astronomy-related doctorates per million population is highest in the least populous countries—Ireland, New Zealand, and Sweden. South Africa and Chile have far fewer degrees per capita than the other countries. This is readily understood since (1) they are poorer countries, and (2) most of their universities did not start awarding doctorates until fairly recently—South Africa in 1972, Chile in 2004. Both have taken advantage of the fact that they provide sites for major optical and radio observatories that are funded and operated by institutions and governments in the rich world. Their astronomers are entitled to some of the observing time at these observatories, and the backers of the observatories have agreed to help them build communities of local astronomers to use this time. In both cases their degree production is increasing rapidly.

A more surprising outlier is California, which with a population just a little (9%) greater than that of Canada has fewer than half as many universities producing nearly three times as many doctorates. However, California is not extreme among the states in any category other than population and total number of degrees: Arizona has produced a whopping 298 doctorates per institution (of which there are only two), and Massachusetts has 171 per million residents.

In terms of doctorates per million population, the Netherlands is the highest among the ten nations, with the U.K. second and the U.S. third, which is not surprising as all three attract a lot of foreign students, while the Netherlands produces the most degrees per institution. The productivity of California institutions is far greater than that of any of the countries, with 222 doctorates per university. It will even be first among the states soon, as Arizona is starting a third doctoral program in astronomy, at Northern Arizona University, and will drop below it. Will China dwarf these numbers? What about universities in other countries? We know that the Department of Physics and Astronomy at the University of Heidelberg currently produces about 100 doctorates per year (Heidelberg, 2016), but we don't yet know how many of these are astronomy-related. It will be better to make such comparisons when we have more countries in our database.

Another interesting fact is how greatly the production of astronomy-related doctorates has increased over time (Figure 2). Although the first

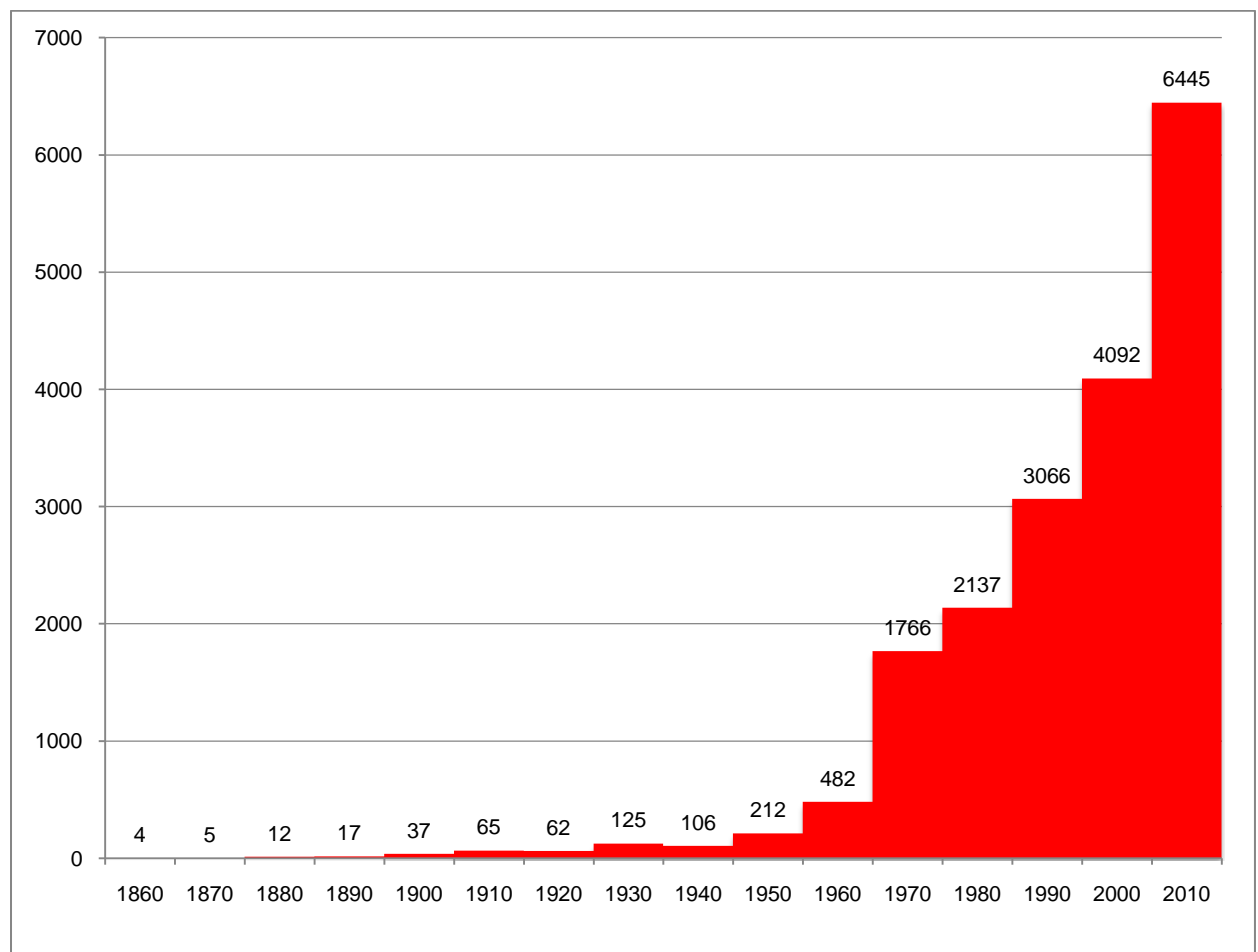

Figure 2: Earned doctorates with astronomy-related theses by decade. The label '1860' means 1856 through 1865, etc. Degrees awarded in 2016 are excluded here (and only here). Data are from the ten countries listed in Table 1. The only decreases were caused by the two World Wars.

U.S. doctorate in astronomy was in 1861, one half of the American degrees have been awarded since 1997. For the other countries listed above, the median year ranges from 1999 (Netherlands) to 2012 (Chile). For the ten countries combined, it is 1999. Of course the Netherlands and the U.K. have been educating astronomers for centuries, but they did not award Ph.D.s in the field until 1863 and 1929 respectively. At least that is the earliest we have found. We would be delighted to learn of earlier such degrees. We are aware of one earlier doctorate: J. Norman Lockyer was awarded a D.Sc. by the University of Cambridge in 1904 for previously-published work. The data for the U.K. are less complete than for the other countries. Some British universities have not yet put their complete catalogues on-line. On the other hand, the University of Manchester's on-line catalogue does not distinguish between Doctoral and Master's degree theses, so we have probably included some of the latter and therefore overcounted

British academics long resisted the Ph.D., considering it an unnecessary foreign invention. Many of the great British astronomers of the twentieth century, including Ralph Howard Fowler (1889–1944), Arthur Stanley Eddington (1882–1944), Fred Hoyle (1915–2001; Figure 3), Hermann Bondi (1919–2005), Martin Ryle (1918–1984), Freeman John Dyson (b. 1923) and Edward Robert Harrison (1919–2007), never earned Ph.D.'s, although some received honorary doctorates late in life.

An excellent example of this attitude is given by Hoyle (1994: 127) in elaborating on his decision to become a student of Rudolph Peierls at Cambridge:

> The situation proved ironic, for, if it had ever been my intention to seek the Ph.D., it was [Maurice] Pryce who persuaded me out of it. He had a dislike for the degree, which he regarded as a debasement of the academic currency. He showed his opinion by fulfilling all the technical requirements but then omitting ever to go to the Senate House to formalize the situation in an official ceremony. As it turned out, I did the same, but only partly for doctrinaire reasons. I discovered the Inland Revenue distinguished between students and nonstudents by whether or not you had acquired the Ph.D., and, since the distinction affected my tax quite substantially in the period 1939–1941, I had a more earthy motive for avoiding an official ceremony in the Senate House.

Hoyle continues with a denunciation of the Ph.D. that includes: "The mere fact that government bureaucracy demands the Ph.D., and has demanded it pretty well from the first moment it was was introduced from America, is sufficient to condemn it." (ibid.)

This illustrates a reason not to keep track of doctoral work only. In fact, AstroGen will eventually include many without doctorates, especially those who were academic ancestors of those who did earn the Ph.D. or equivalent. We already have some. Otherwise most family trees would be short, and none would go back beyond the nineteenth century.

The above information was easily obtained from the spreadsheets. Those historians willing to spend more time with the database will be able to learn how subjects, such as X-ray astronomy, planetary exploration, exoplanets, or gravitational wave astronomy, grew in popularity over time while celestial mechanics and astrometry declined. (The latter has enjoyed a resurgence in recent years with the European Space Agency's *Hipparcos* and *Gaia* missions.) They will also be able to compile information as to the careers of Ph.D. astronomers. For example, in the twenty-first century the number of astronomers using their expertise with 'big data' in such fields as internet companies, financial institutions and retailers may exceed the number who have obtained research positions in physical science. This is certainly true for the graduates of some universities.

It is also possible to compare universities. While 264 universities in the ten countries have awarded a total of 18,923 doctorates with astronomy-related theses, more than three-eighths of these have come from the seventeen largest producers, listed in Table 2. These are the universities that have produced more than 300 doctorates each.

We have not compiled any quantitative information about the careers of those who have earned doctorates with astronomy-related dissertations, but I can make a few comments from reading thousands of acknowledgements in theses and finding and reading current web pages.

In the 1970s and 1980s it was almost unthinkable for a graduate student to express religiosity in a thesis. It has become increasingly common in recent years, both in the United States and in some parts of Europe. Although by no means a large percentage of theses, there are now many in which the author expresses his or her religious views, sometimes at considerable length. This is always done in the acknowledgements section.

Individuals who survive graduate school in the physical sciences are an enterprising lot. As the fraction who obtain research positions in the field has decreased, graduates have made careers in many fields that one might not expect, as doctors, lawyers, entertainers, fiction and non-fiction writers, clergy of all faiths, public speakers, photographers, entrepreneurs, and in many other areas. Of course most have found ways to use their educations more directly. Those whose degrees are in electrical engineering, computer science or the earth sciences usually work in those areas. Many astronomy graduates go into defense industries and government laboratories. Some teach in secondary school or colleges. Some make their livings as communicators of science to the public. And, as mentioned above, many use their skills at manip-ulating 'big data' in fields such as finance.

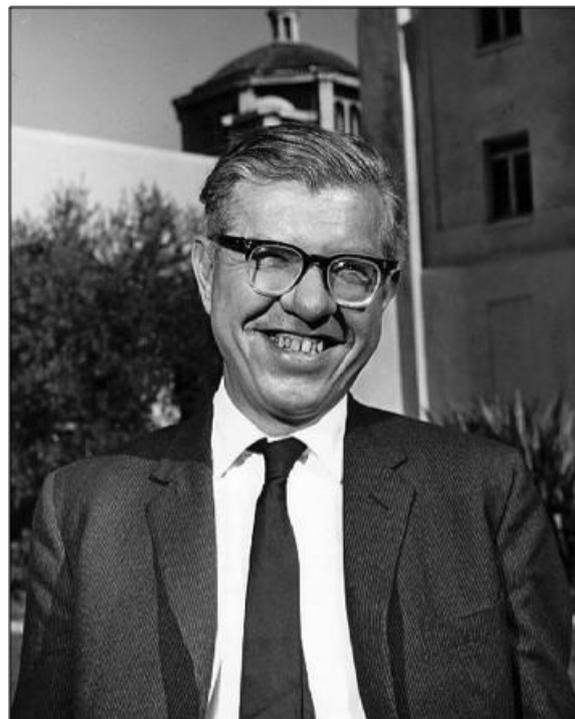

Figure 3: Fred Hoyle at Caltech in 1967 (Courtesy: Clemson University and Donald D. Clayton).

Table 2: The seventeen universities in our ten countries that have produced 300 or more Ph.D.'s with astronomy-related theses as of November 2016. This is slightly more up-to-date for some universities than for others.

| University | Doctorates |
| --- | --- |
| University of California, Berkeley | 652 |
| California Institute of Technology | 565 |
| University of Cambridge | 536 |
| University of Arizona | 512 |
| Harvard University | 496 |
| University of Chicago | 458 |
| University of Texas at Austin | 412 |
| University of Maryland, College Park | 399 |
| Princeton University | 389 |
| University of Colorado Boulder | 379 |
| Massachusetts Institute of Technology | 366 |
| Cornell University | 342 |
| University of Leiden | 333 |
| University of Wisconsin-Madison | 321 |
| University of Michigan | 320 |
| University of Manchester | 318 |
| University of California, Los Angeles | 316 |

Another qualitative observation is that ethnicity and gender matter. A graduate student with an Asian, Hispanic, Middle Eastern or Slavic name is more likely to choose a thesis advisor with the same background than would be expected by chance, and a female student is much more likely to choose a female professor. This observation is difficult to quantify, as we would need to know the ethnic and gender distributions of the department at the time of the thesis.

## 5 WHAT NEXT?

It may take a while before the programmers have completed the necessary work to convert our spreadsheets to a polished website. In the meantime we are anxious to correct errors (there are certain to be some in what we have gathered from the web), add more information, such as the names of advisors of those whose theses are not on-line, and, especially, expand from ten countries to the world. (Since submitting the first version of this paper we have completed Norway and a good portion of Spain.) We are seeking volunteers. If you know the language and something of the academic culture of another country, we would very much like to have you join us and gather information on some of the theses from that country. If you would like to work on our list of universities, that would also be helpful. If you can go into a university library, get old theses out of storage, and photograph the page listing the advisor and the acknowledgements section or copy the necessary information, then you could make a major contribution to the Project. We also welcome comments and suggestions. Please contact me at astrogendirector@aas.org.

## 6 ACKNOWLEDGEMENTS

Those who have compiled information for AstroGen include Jennifer Bartlett, Sally Bosken, Peter Broughton, John Gerard Doyle, Andrew Fabian, David J. Helfand, Mark Hurn, Matthew Knight, James Lattis, Warrick Lawson, Jordan Marché, Eugene Milone, Donald C. Morton, Ryan Quitzow-James, Katherine Rhode, Gordon Robertson, Arnold Rots, Patrick Seitzer, Horace Smith and the author. We also appreciate valuable advice provided by Mitch Keller, the director of the MGP, and by Richard Jarrell (1946–2013), Marc Rothenberg, and several others.

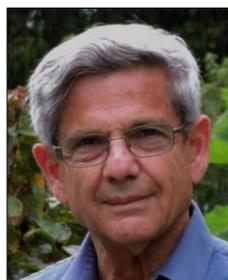
Joseph S. Tenn taught physics and astronomy at Sonoma State University in the California wine country from 1970 to 2009. He served as Secretary-Treasurer of the Historical Astronomy Division of the American Astronomical So-ciety from 2007 to 2015 and as an Associate Editor of the *JAHH* from 2008 to 2016. He maintains the Bruce Medalists website at http://phys-astro.sonoma.edu/brucemedalists/. His main current activity is directing the Astronomy Genealogy Project. His ORCID number is 0000-0002-7803-3633.

## POSTSCRIPT

The author has very recently discovered several hundred more doctorates, mostly in the UK, which he is currently adding to the database. While it is too late to update this paper, note that University College London will join the list of universities that have awarded more than 300 astronomy-related doctorates.